\documentclass[12pt]{article}
\usepackage[T2A]{fontenc}
\usepackage[cp1251]{inputenc}
\usepackage{a4wide}
\usepackage{amsmath,latexsym,amsthm,amsfonts}
\usepackage{amssymb,amscd}
\usepackage{graphicx}

\newcommand{\bml}[1]{\begin{multline}\label{#1}}
\newcommand{\bee}{\begin{equation}}
\newcommand{\bed}{\begin{displaymath}}
\newcommand{\ee}{\end{equation}}

\newcommand{\bs}{\begin{split}}

\newcommand{\ga}{\gamma} \newcommand{\Ga}{\Gamma}
 \newcommand{\De}{\Delta}
\newcommand{\la}{\lambda}

 \newcommand{\sm}{\setminus}

\newtheorem{theorem}{Theorem}[section]

\newtheorem{remark}{Remark}[section]
\newtheorem{corollary}{Corollary}[section]
\theoremstyle{definition} 

\theoremstyle{remark} 
 \numberwithin{equation}{section}
\begin{document}

\title{On  stability, superstability and strong superstability of classical systems of Statistical Mechanics.}
\author{A.~L.~Rebenko$^1$, M.V.~Tertychnyi $^2$}
\date{}
\maketitle
\begin{footnotesize}
\begin{tabbing}
$^1$ \= Institute of Mathematics, Ukrainian National Academy of Sciences, Kyiv, Ukraine\\
\> {\em rebenko@voliacable.com  ; rebenko@imath.kiev.ua}\\
$^2$ \> Faculty of physics, Kyiv Shevchenko university, Kyiv, Ukraine \\
\>{\em mt4@ukr.net}\\
\end{tabbing}
\end{footnotesize}
\begin{abstract}
A detailed analysis of conditions on 2-body interaction potential,
which ensure stability, superstability or strong superstability of
 statistical systems is given. There has been given
the connection between conditions of superstability (strong
superstability) and the problem of minimization of  Riesz energy
in the bounded volumes.
\end{abstract}
\thispagestyle{empty}

\noindent \textbf{Keywords :} Continuous classical system;
superstable interaction; minimal Riesz energy.\\

\noindent \textbf{Mathematics Subject Classification :}\, 82B05;
 82B21

\section{\bf Introduction}
Stability\,{\bf (S)} of the interaction is  a necessary condition
for the  correct thermodynamic description of infinite statistical
systems. This condition can be formulated by infinite system of
inequalities on the interaction energy of an arbitrary finite
subsystem, consisting of $N$ particles, which are situated in the
points $x_{1},...,x_{N}$ of the space $\mathbb{R}^{d}$.

{\bf (S)}\textit{Stability. There exists $B\geq 0$ such that
\begin{equation}\label{E:S}%(1.1)
  U(x_{1},...,x_{N})\geq -B\,N
\end{equation}
for any $N\geq 2$ and $\{x_{1},...,x_{N}\}$}.

 In the present paper we consider an infinite system, which
 consists of identical point particles interacting via 2-body potential
\begin{equation}\label{potential}%(1.2)
  V_{2}(x,y)=\Phi(|x-y|),
\end{equation}
where $|x-y|$ means Euclidean distance between points $x,y\in
\mathbb{R}^{d}$. In this case
\begin{equation}\label{E:interaction}%(1.3)
U(x_{1},...,x_{N})=\sum_{1\leq i<j \leq N} \Phi(|x-y|).
\end{equation}
One of the most important conditions is the \textit{condition of
integrability at the infinity}. This means that for any \, $R>0$\
\begin{equation}\label{E:integrability}%(1.4)
 \int_{|x|\geq R}\Phi (|x|)\,dx < +\infty.
\end{equation}
The conditions\, (1.1)\, and\, (1.4) are sufficient for the
construction of Gibbs measure of an infinite system of particles
in the area of small values of parameters $\beta = \frac{1}{k_BT}$
\;and $z$,\, where $T$ is a temperature of a system and
 $z$ is a chemical activity, which is directly connected with a
density of the system of particles(see for example [25], ch.4). In
order to solve the problem of construction of Gibbs state(Gibbs
measure) of an infinite system for all positive values of
parameters $\beta$ and $z$, it is necessary to impose more
restrictive conditions on the interaction. Such a condition is the
condition of superstability\,{\bf (SS)}(see [8],
[26]). At first we give several necessary definitions.\\
For each $\lambda\in\mathbb{R}_{+}$ one can  define the partition
$\overline{\Delta_{\lambda}}$\,of the space $\mathbb{R}^{d}$ into
cubes $\Delta$ with a rib $\lambda$ and center in
$r\in\mathbb{Z}^d$:
\begin{equation}\label{E:part}%(1.5)
\Delta =\Delta_{\lambda}(r):=\left\{ x\in \mathbb{R}^{d}\; \mid
\;\lambda\left(r^{i}-1/2 \right)\leq x^{i}< \lambda \left( r^{i}+1/2
\right)\right\}.
\end{equation}
Let\, $\Gamma$\, be a phase space of an infinite statistical system
of identical point particles. In the  case of an equilibrium system
\, $\Gamma$\, coincides with the space of
\textit{configurations}\,(in our situation coordinates of
particles)\, $\gamma$\, which are locally finite subsets of
$\mathbb{R}^{d}$. In other words
\begin{equation}\label{E:finconf}%(1.6)
 \Gamma:=\left\{\gamma \subset \mathbb{R}^{d}\,| \,|\gamma \cap
\Lambda |< \infty,\,\textnormal{for all}\, \Lambda \in
\mathcal{B}_c(\mathbb{R}^{d})\right\},
\end{equation}
where $\mathcal{B}_c(\mathbb{R}^{d})$ is a set of all bounded Borel
subsets of $\mathbb{R}^{d}$,\,and  $|X|$\, is the cardinality of a
set $X\Subset \mathbb{R}^{d}$. Let us define also the subset\,
$\Gamma_{0}$\, of all finite configurations:
\begin{equation}\label{E:conf}%(1.7)
  \Gamma_{0}= \underset{n\in \mathbb{N}_{0}}{\coprod}\Gamma^{(n)}\,
  , \quad \Gamma^{(n)}:=\left \{\gamma\in \Ga \; | \;
  |\gamma|=n \right \},\; \mathbb{N}_{0}=\mathbb{N}\cup\{0\}.
\end{equation}
Besides, let
\begin{equation}\label{E:projection}%(1.8)
\gamma_{\Lambda}:=\gamma\cap\Lambda,\,
\gamma\in\Gamma,\,\Lambda\in\mathcal{B}_c(\mathbb{R}^{d}).
\end{equation}
{\bf (SS)} \textit{Superstability.\, There exist $A>0, \,B\geq0 $
and the partition $\overline{\Delta_{\lambda}}$ such that for any
$\gamma=\{x_{1},\ldots,x_{N}\}\in \Gamma_{0}$ the following holds:}
\begin{equation}\label{E:SS}%(1.9)
 U(\gamma)\geq  \underset{\Delta\in \overline{\Delta_{\lambda}} }
{\sum}\;\left[ A |\gamma_{\Delta}|^2 - B|\gamma_{\Delta}| \right ].
\end{equation}
\begin{remark}\label{R:Ginibre}
A slightly different definition was introduced by Ginibre (see [8]
):
\\An interaction is superstable if there exist two real
constants\, $B\geq 0$\, and $A_{1}\geq0$\, such that for any
$\gamma\in\Gamma_{0}$ the following is true:
\begin{equation}\label{E:SS'}%(1.10)
U(\gamma)\geq A_{1}\frac{|\gamma|^2}{\xi^{d}}-B|\gamma|,
\end{equation}
where $\xi=\underset{\{x,y\}\subset \gamma}{max}|x-y|$. Let us
consider  a box $\Lambda$\, with a volume $V=vol\,(\Lambda)$\,
such that $\gamma\subset\Lambda$. Then the condition (1.10) can be
rewritten in the following form:
\begin{equation}\label{E:SS''}%(1.11)
U(\gamma)\geq A_{\Lambda}\frac{|\gamma|^2}{V}-B|\gamma|,
\end{equation}
where the constant $A_{\Lambda}$\, does not depend on the volume
$V$ for the given shape, but it may be shape dependent. It is easy
to notice, that if we consider the box $\Lambda$\, as a union of
the cubes $\Delta$,\,defined by (1.5)\,and containing at least one
point of the configuration $\gamma$,\,then, using Cauchy-Schwarz
inequality, one can write the following inequality:
 \[
|\gamma|^{2}=\left( \sum_{\Delta\in
\overline{\Delta_{\lambda}}}|\gamma_{\Delta}| \right)^{2}\leq
\sum_{\Delta \in \overline{\Delta_{\lambda}}\cap
\gamma}\,1\cdot\sum_{\Delta\in
\overline{\Delta_{\lambda}}}|\gamma_{\Delta}|^{2}=\frac{V}{\lambda^{d}}
\sum_{\Delta\in\overline{\Delta_{\lambda}}}|\gamma_{\Delta}|^{2}.
\]
So, the condition (1.11) follows directly from\, (1.9) with
$A_\Lambda=A\lambda^d$.
\end{remark}
There is a stronger condition on the interaction than (1.9).

{\bf (SSS)} \textit{Strong superstability.\, There exist $A>0,
\,B\geq 0,\, p\geq 2 $ and the partition
$\overline{\Delta_{\lambda_0}}$ such that for any
$\gamma=\{x_{1},\ldots,x_{N}\}\in \Gamma_{0}$ the following holds:}
\begin{equation}\label{E:SSS}%(1.12)
 U(\gamma)\geq  \underset{\Delta\in \overline{\Delta_{\lambda}} }
{\sum}\;\left[ A |\gamma_{\Delta}|^p - B|\gamma_{\Delta}| \right ].
\end{equation}
{\it for any $\lambda\leq\la_0$}

 V.~M.~Park (see [19]) was
the first, who used the condition (1.12) with $p>2$ for the proof
of bounds on exponentials of local number operators
of quantum systems of interacting Bose gas.\\
In connection with the conditions\, (1.1),\, (1.9),\, (1.12) there
is a problem to describe the behavior of interaction potentials,
which ensure the stability, superstability or strong
superstability of the statistical systems. Putting in \,
(1.1)\,-\,(1.3)\, $N=2$, we deduce that the function $\Phi$ must
be bounded from below:
\begin{equation}\label{E:lowbound}%(1.13)
  \Phi(|x|)\geq -2B.
\end{equation}
In addition to this,  R.L.Dobrushin (see [6]) proposed  a
\textit{necessary} condition of stability \\of interaction in the
form:
\begin{equation}\label{E:nesstab}%(1.14)
\int_{\mathbb{R}^{d}}\Phi(|x|)\,dx\geq 0.
\end{equation}
Consequently, a positive part of interaction must be big enough. As
a rule, for neutral physical systems, the potential with the
behavior as on the Figure 1 is considered.
\begin{align}
& \includegraphics{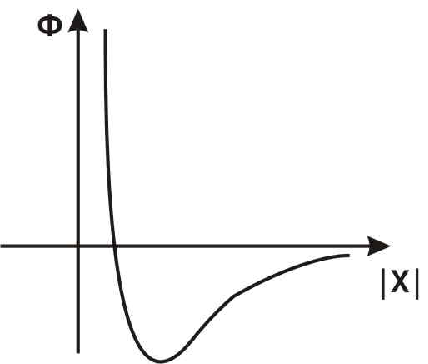} \notag\\
& \textnormal{Fig. 1} \notag
\end{align}
A behavior of the potentials at the infinity\,
$(|x|\rightarrow\infty)$\,is determined by the condition (1.4),
but the behavior near the initial point depends on the chosen
model and as we will see later, it actually defines {\bf
(S),\,(SS),\,(SSS)}\, type of  interaction. D.~Ruelle was the
first, who introduced the conditions, which ensure the estimate
(1.11)\,  for the systems of particles, which are situated in the
cube \,$\Lambda$\, with a volume \,$V$\,(see [24]). He proposed
the potential\, $\Phi$\,  in the following form:
\begin{equation}\label{E:form}%(1.15)
\Phi(|x|)=\Phi_{1}(|x|)+\Phi_{2}(|x|),
\end{equation}
where\, $\Phi_{1}$\, is Lebesgue measurable function with values
in the closed interval $[0;\infty ]$ and satisfies the condition
(1.4);\, $\Phi_{2}$\, is a continuous  function of positive type
and:
\begin{equation}\label{E:postype}%(1.16)
  \overset{\sim}{\Phi_{2}}(0)=\int_{\mathbb{R}^{d}}\,\Phi_{2}(x)\,dx>0.
\end{equation}
The above mentioned conditions and their direct consequence - the
inequality \, (1.11)\,were used in works  [24]  for the proof of
existence of a thermodynamic limit\,$(\Lambda\nearrow
\mathbb{R}^{d})$\, for a free \\
energy\,(canonical ensemble) and a pressure (grand canonical
ensemble). Later M.~Fisher noticed(see remarks in [24])\; that
these results can be proved using less restrictive assumptions on
the potential\, $\Phi$:
\begin{align}
&\Phi(|x|)\geq \frac{c}{|x|^{d+\varepsilon}}\;  \text{for}\;
|x|<a_{1},\label{E:cond1}\\%(1.17)
&\Phi(|x|)\geq -w\; \text{for}\;  a_{1}\leq |x|\leq a_{2},\label{E:cond2}\\%(1.18)
&\Phi(|x|)\geq - \frac{c'}{|x|^{d+\varepsilon'}}\; \text{for}\;
|x|>a_{2},\label{E:cond3}%(1.19)
\end{align}
where $a_{1},\,a_{2},\,c,\, c',\, w,\,\varepsilon,\,
\varepsilon'$\, are some positive constants. See also the article
[7] for the systems of particles with different species and
"charged" systems. As the authors pointed out,the conditions\,
(1.17)\,-\,(1.19)\, ensure {\bf (S)} stability of a system, in
other words the condition \, (1.1)\, holds. In fact, these
conditions guarantee also superstability of interaction.
But at that time such a notion was not yet introduced.\\
Independently, and at the same time A.Ya. Povzner(communication at
the Moscow State University seminar on Statistical Mechanics
(1963)) found  the conditions on the potential, which ensure the
existence of the estimate \,(1.1)(and even (1.9)). One can find
his arguments in \, [28] where they have been refined for the
analysis of stability of the classical statistical systems with
highly singular potentials. Later R.L.Dobrushin proposed more
general condition on the potential $\Phi$, which in contrast to
(1.17) included also integrable at the origin potentials\,(see
[6], formula\,\eqref{E:cond1}). Having modified Povzner conditions
he proved, that stability and an existence of limit values of
thermodynamic potentials follow from these conditions. In order to
complete this short survey we have to mention the criterion of
stability, which was proposed by Basuev [2]. Note that it is
rather close to the Povzner's conditions (see also
[20]).\\
In terms of usage of the conditions \, (1.9),\, (1.11)\, it is
important to obtain the optimal values of the constants\,
$A,\,B$.\, In this area we have to mention the article \,[17]\, in
which for continuous\, $L^{1}(\mathbb{R}^{d})$\, potentials of
positive type, which satisfy the condition (1.16),\, the
inequality\, (1.11)\, was proved with
\[
  A=\frac{1}{2}\left(
  \overset{\sim}{\Phi}(0)-\varepsilon\right),\,B=\frac{1}{2}\Phi(0),\,
  \text{and}\; V=V(\varepsilon)
\]
for any small\, $\varepsilon>0$.\, The constants \,$A,B$\, are
best possible.\\
The purpose of the present article is not only to make an overview
of the previous results, but to obtain some new sufficient
conditions on the 2-body interaction potential, which make a
system stable, superstable or strong superstable. It is important
to notice that the remark about the possible behavior of singular
potentials, which ensures the condition\, (1.12)\, for\, $p>2$\,
was firstly proposed by D.Ruelle\,(see [25], ch.3, formula(2.28)).
 It seems to be just an intuitive assumption, which one can
guess on the physical level of rigor,\,if we accept the following
hypothesis: the configuration that minimizes energy of\,
$N$\,particles, which are situated in the cube with a volume
\,$V$\,is uniformly distributed. It means, that all particles are
situated in the sites of a lattice on the
distances\,$\sim\left(\frac{V}{N}\right)^{\frac{1}{d}}$.\,
Implicitly such estimate of the energy was calculated also by
Dobrushin (see [5] formulas (4.1), (3.2)). Therefore, the present
work can be considered as a new proof of Ruelle's conjecture [25].
We used rigorous results, that have been obtained during last
several years(see [3], [9], [10], [11] ) and some facts of the
classical potential theory (see, for example [13]). Besides, exact
values of the constants $A$ and $B$ in the Eqs. (1.9), (1.12) are
established.

\section{\bf Notations and main results}

 Following [13]\;let us propose several new
 notations, some of them will be denoted in accordance with
 the chapter 1 of the present article. Let\,
 $K$ be a compact in $\mathbb{R}^{d}$.\,For any
 configuration\,$\gamma_{K}(|\gamma_{K}|=N)$\, in $K$\, we define
 the Riesz s-energy:
\begin{equation}\label{E:Riesz}%(2.1)
  E_{s}^{(N)}(\gamma_{K}):=\sum_{\{x,y\}\subset
  \gamma_{K}}\frac{1}{|x-y|^{s}},\, s>0.
\end{equation}
In the case \,$s<d$\, consider\, \textit{the energy integral}
\begin{equation}\label{E:energint}%(2.2)
  I_{s}(\mu;K):=\frac{1}{2}
  \underset{K \times K}{\int\int}\frac{1}{|x-y|^{s}}\,\mu(dx)\mu(dy),
\end{equation}
w. r. t. some probability  measure \,$\mu$,\,support of
which is \,$K$\, ( $\mu(K)=1$).\, \\
One of the most important problems in  modern potential theory is
to find a  measure\,$\mu^{\ast}$\,that minimizes the integral\
(2.2).\\ There is a fact\,(see [13], ch. 2)\, that if the
configuration \,$\gamma_{K}^{min}=\{\xi_{1},\ldots,\xi_{N}\}$\,
minimizes\\
the energy\,\eqref{E:energint},\;then a sequence of measures:
\begin{equation}\label{E:seqmeasure}%(2.3)
  \mu_{N}(\cdot):=\frac{1}{N}\sum_{i=1}^{N}\delta_{\xi_{i}}(\cdot),
\end{equation}
where\, $\delta_{\xi_{i}}$\, is a point Dirac measure, converges
in the weak-star topology to the
measure\,$\mu^{\ast}$\,(minimizing measure of the integral\, (2.2)).\\
A
sequence\,$e_{s,K}^{(N)}=\frac{E_{s}^{(N)}(\gamma_{K}^{min})}{N^{2}}$\,
is monotonically increasing and:
\begin{equation}\label{E:incseq}%(2.4)
\underset{N\rightarrow\infty}{\textnormal{lim}}e_{s,K}^{(N)}=\underset{N\rightarrow\infty}{\textnormal{lim}}
\frac{E_{s}^{(N)}(\gamma_{K}^{min})}{N^{2}}=I_{s}(\mu^{\ast})<\infty.
\end{equation}
There are two different behaviors of the minimizing configurations
in the
limit $N\rightarrow\infty$:\\
 1)if $s\leq d-2$,\,\,then \;$\textnormal{supp}\,
\mu_{N}\subset\partial K,\,\textnormal{supp}\,
\mu^{\ast}\subset\partial K $,\, where $\partial K$ is a border of a
compact $K$;\,2) for \,  $d-2<s< d \,\, \textnormal{supp}\,
\mu^{\ast}\subset K$.

 Let $K=\mathcal{B}^{d}(0;r)$\, be a
d-dimensional ball with a radius $r$\, and $\partial K=S^{d}(0;r)$\,
be a surface of the corresponding sphere. Than for the  case  1)\,
for $s\leq d-2$ \, minimizing measure is distributed uniformly on
the surface of the ball\, $\mathcal{B}^{d}(0;r)$\, and:
\begin{equation}\label{E:firstmeasure}%(2.5)
  \mu^{\ast}(dx;\mathcal{B}^{d}(0;r))=
  \frac{m(dx)|_{S^{d}(0;r)}}{m(S^{d}(0;r))},\,
  m(S^{d}(0;r))=\frac{2\pi^{\frac{d}{2}}}{\Gamma\left(
  \frac{d}{2}\right)}r^{d-1};
\end{equation}
2)\, for  $d-2<s< d$
\begin{equation}\label{E:secondmeasure}%(2.6)
  \mu^{\ast}(dx;\mathcal{B}^{d}(0;r)=\frac{A(d;s)}{(r^{2}-x^{2})^{\frac{d-s}{2}}}m(dx),\,
  A(d;s)=\frac{\Gamma\left(1+\frac{s}{2}\right)}{\pi^{\frac{d}{2}}\Gamma\left(1-\frac{d-s}{2}\right)},
\end{equation}
where \,$m(\cdot)$\, is  the Lebesgue measure in $\mathbb{R}^{d}$.\,
Corresponding values of the energy integral\,
(2.2) are:\\
1)\, for $s\leq d-2$
\begin{equation}\label{E:firstint}%(2.7)
I_{s}(\mu^{\ast};\mathcal{B}^{d}(0;r))=\frac{1}{r^{s}}\,\frac{2^{d-s-3}\;\Gamma\left(\frac{d-s-1}{2}
\right)\Gamma\left(\frac{d}{2}\right)}{\sqrt{\pi}\;\Gamma
\left(d-1-\frac{s}{2}\right)},
\end{equation}
2)\, if $d-2<s< d$
\begin{equation}\label{E:secondint}%(2.8)
I_{s}(\mu^{\ast};\mathcal{B}^{d}(0;r))=\frac{1}{r^{s}}\,
\frac{\Gamma\left(1+\frac{s}{2}\right)\;\Gamma\left(\frac{d-s}{2}\right)}{2\;\Gamma\left(1+\frac{d}{2}
\right)}.
\end{equation}
See for details \, [13].\\

The cases $s=d$ and $s > d$ are essentially different from the
case $s<d$, which is considered in the classical potential theory.
The construction of the minimizing measure and the estimates for
the minimal energy of the configuration if $s\geq d$ are proposed
in [9] -
[11] (see, also, [3]).\,Let us formulate the most important points:\\
1) the energy integral $I_{s}(\mu)=+\infty$ for all probability
measures on the compact $K\subset \mathbb{R}^{d}$; \\
2) for any arbitrary compact $K\subset \mathbb{R}^{d}$ the following
is true:
\begin{equation}\label{E:uniformdist}%(2.9)
 \mu_{N}(\cdot)\rightarrow \frac{m(\cdot)|_{K}}{m(K)}.
\end{equation}
or in other words point particles are asymptotically uniformly
distributed;\\
3) if $s=d$ the following holds:
\begin{equation}\label{E:s=d_case}%(2.10)
 C_{d}=\underset{N\rightarrow \infty}{lim}\frac{E_{s}^{N}\left(\gamma_{K}^{min}\right)}{N^{2}\ln N}=
 \frac{\varphi_{0}}{\lambda^{s}}
 \frac{\pi^{\frac{d}{2}}}{d\cdot\Gamma\left( \frac{d}{2}\right)};
\end{equation}
4) if $s>d$ then:
\begin{equation}\label{E:s>d_case}%(2.11)
\underset{N\rightarrow\infty}{lim}\frac{E_{s}^{N}\left(\gamma_{K}^{min}\right)}{N^{1+\frac{s}{d}}}=
\frac{\varphi_{0}}{\lambda^{s}}\frac{C_{s,d}}{2}.
\end{equation}
In the case $d=1$ and $K\,=\,[0,1]$\; $C_{s,1}=2\xi(s)$, where $\xi(s)$  is a classical Riemann zeta-function; \\
5) let $K$ be a d-dimensional cube with a rib\, $\lambda$, then if
$s>d$ the following holds:
\begin{equation}\label{E:s>d_case_add}%(2.12)
E_{s}^{N}\left(\gamma_{K}\right)\geq
\frac{\varphi_{0}}{\lambda^{s}}\frac{1}{2^{2s+1}}\left(
\frac{2\pi^{\frac{d}{2}}}{d\cdot
\Gamma\left(\frac{d}{2}\right)}\right)^{\frac{s}{d}}N^{1+\frac{s}{d}}.
\end{equation}

{\bf (A): Assumption on the interaction potential.} \textit{In this
article we consider a general type of potentials \,$\Phi$,\, which
are continuous on $\mathbb{R_+}\sm\{0\}$, and for which there exists
\,$\lambda> 0,\,R>\lambda,\,\varphi_{0}>0,\,\varphi_{1}>0,\,
\textnormal{and}\, \epsilon>0$\, such that}:
\begin{align}
&1)\, \Phi(|x|)\equiv\Phi^-(|x|)\geq -
\frac{\varphi_{1}}{|x|^{d+\epsilon}}\;
  \textnormal{for}\; |x|\geq R,\label{E:fincond1};\\%(2.13)
&2)\, \Phi(|x|)\equiv\Phi^+(|x|)\geq \frac{\varphi_{0}}{|x|^{s}},\,
s\geq 0\;
\textnormal{for}\, |x|\leq \lambda.\label{E:fincond2}%(2.14)
\end{align}
\textit{where}
\begin{equation}\label{214} %(2.15)
\Phi^+(|x|):= \max \{0, \Phi(|x|)\}, \, \Phi^-(|x|):=\min \{0,
\Phi(|x|)\}.
\end{equation}
In contrast to\, [9],\, [13]\, we consider also the case $s=0$,
which looks probably trivial from the point of view of
 potential theory , but it will take place also in our
description(see Remark 2.1 below).\\
Now we can formulate the following theorems.
\begin{theorem}\label{T:first}
Let interaction potential satisfy the conditions {\bf (A)}. Then for
$0\leq s<d$\, any $\gamma\in\Gamma_{0}$\, and  sufficiently small\,
$\varepsilon>0$\, there exists constant $B=B(\varepsilon)$ such that
the following inequality holds :
\begin{equation}\label{E:t1}%(2.16)
U(\gamma)\geq
\underset{\substack{\Delta\in\overline{\Delta_{\lambda}},\\|\gamma_{\Delta}|\geq
2 }}{\sum}\left(I_{s}(\mu^{\ast};\Delta)\varphi_{0}-\frac{v_0}{2}
-\varepsilon\right)|\gamma_{\Delta}|^{2}- B|\gamma|,
\end{equation}
where
\begin{equation}\label{E:t1_not}%(2.17)
v_{0}=v_{0}(\lambda):=\sup_{x\in\mathbb{R}^d}\sum_{\Delta\in
\overline{\Delta_{\lambda}}}\underset{y\in
\Delta}{\;\;\textnormal{sup}}\left|\Phi^{-}(|x-y|)\right|.
\end{equation}
\end{theorem}
\begin{corollary}\label{C:t1}
In the case: \,$0\leq s<d$\, the potential $\Phi$\, yields the
condition {\bf (SS)}(see (1.9)\, if the following holds:
\begin{equation}\label{E:ss_cond}%(2.18)
  I_{s}(\mu^{\ast};\Delta)\varphi_{0}> \frac{v_0}{2}.
\end{equation}
\end{corollary}
\begin{remark}\label{R:addtot1}
The condition \,(2.18)\, can be rewritten in simpler form if we
consider that minimal Riesz energy of a configuration with fixed
number of particles $|\gamma_{\Delta}|$\, in the cube
$\Delta\in\overline{\Delta_{\lambda}}$ is always bigger than
minimal Riesz energy of a configuration with the same number of
particles in the described ball with a radius
$r=\frac{\sqrt{d}\lambda}{2}$.\, Consequently, one can substitute
the formulas (2.7),\, (2.8)\, with $r=\frac{\sqrt{d}\lambda}{2}$\,
for $I_{s}(\mu^{\ast};\Delta)$\,in the l.h.s of (2.18)\,(cases:
$s\leq d-2,\,d-2<s<d$\, respectively). The r.h.s of (2.18) can be
changed by $\frac{C}{\lambda^{d}}$,\, where a constant $C\approx
\int_{\mathbb{R}^{d}}\left| \Phi^{-}(|x|)\right|\,dx$ for
sufficiently small $\lambda$.\,Then for the given configuration in
the d-dimensional space and for the potential, which satisfies the
condition (2.14)\, the system is superstable if there exists such
$\lambda$\,(in other words such a partition
$\overline{\Delta_{\lambda}}$\ of the space $\mathbb{R}^{d}$),
that the condition (2.18)\, holds. The set of potentials, which
satisfy the condition {\bf (SS)}, is not empty, as  one can choose
sufficiently big $\varphi_{0}$, in order to make (2.18) true for
any fixed $\lambda>0$. For the case $s=0$
\;$I_{0}(\mu^{\ast};\Delta)= 1/2$.
\end{remark}
\begin{theorem}\label{T:second}
Let interaction potential satisfy conditions {\bf (A)}. Then
for\,\;$s=d$, any $\gamma\in\Gamma_{0}$\, and  sufficiently small\,
$\varepsilon>0$\;there exists constant $B=B(\varepsilon)$ such that
the following inequality holds :
\begin{equation}\label{E:t2}%(2.19)
U(\gamma)\geq
\underset{\substack{\Delta\in\overline{\Delta_{\lambda}},\\|\gamma_{\Delta}|\geq
2
}}{\sum}\left(C_{d}\,\ln\,|\gamma_{\Delta}|-\frac{v_0}{2}-\varepsilon
\,\ln\,|\gamma_{\Delta}| \right)|\gamma_{\Delta}|^{2}- B|\gamma|,
\end{equation}
where (see [9])
\begin{equation}\label{E:const1}%(2.20)
C_{d}=\frac{1}{\lambda^{d}}\frac{\pi^{\frac{d}{2}}}{d\,\Gamma\left(\frac{d}{2}\right)}\,\varphi_{0}.
\end{equation}
\end{theorem}
\begin{theorem}\label{T:third}
Let interaction potential satisfy conditions {\bf (A)}. Then for\;\;
$s>d$\, any $\gamma\in\Gamma_{0}$\,there exists constant
$B=B(\varepsilon)$ such that the following inequality holds:
\begin{equation}\label{E:t3}%(2.21)
U(\gamma)\geq
\underset{\substack{\Delta\in\overline{\Delta_{\lambda}},\\|\gamma_{\Delta}|\geq
2
}}{\sum}\left(C_{s,d}\,|\gamma_{\Delta}|^{1+\frac{s}{d}}-\frac{v_0}{2}|\gamma_{\Delta}|^{2}\right)-B|\gamma|,
\end{equation}
where (see [9])
\begin{equation}\label{E:const2}%(2.22)
C_{s,d}=\frac{1}{\lambda^{s}}\frac{1}{2^{2s+1}}\left(\frac{2\pi^{\frac{d}{2}}}{d\,\Gamma\left(\frac{d}{2}\right)}\right)^{\frac{s}{d}}\varphi_{0}.
\end{equation}
\end{theorem}
\begin{remark}\label{R:final}
In the case $s=d$\, the system of particles is superstable {\bf
(SS)}\, for all partitions $\overline{\Delta_{\lambda}}$\, since for
any $\varepsilon>0$ and $v_0$ one can find $N_0\geq 2$ and
$B=B(N_0)$ such that for $N>N_0$
\begin{equation}\label{E:rem2}%(2.23)
C_{d}\,\ln\,N> \frac{v_0}{2}.
\end{equation}
In the case\, $s>d$\, system of particles is strong superstable
{\bf (SSS)}, since one can always choose sufficiently small\,
$\lambda>0$\, and some $A=A(\lambda)$\, such that:
\begin{equation}\label{E:SSSrem}%(2.24)
C_{s,d}|\gamma_{\Delta}|^{1+\frac{s}{d}}-\frac{v_{0}}{2}|\gamma_{\Delta}|^{2}\geq
A|\gamma_{\Delta}|^{1+\frac{s}{d}}
\end{equation}
for $|\gamma_{\Delta}|\geq2$.
\end{remark}
\section{\bf Proof of the results}

\subsection{\bf Proof of Theorem 2.1}

We have for any \,$\gamma\in\Gamma_{0}$\, and any partition\,
$\overline{\Delta_{\lambda}}$:

\begin{equation}\label{E:p_t1_1}%(3.1)
U(\gamma)=\sum_{\{x,y\}\subset\,\gamma}\Phi(|x-y|)=\sum_{\Delta\in\,\overline\Delta_\lambda:|\ga_\De|\geq
2}U(\gamma_\De)
+\sum_{\{\Delta,\Delta'\}\subset\overline\Delta_\lambda}\sum_{\substack {x\in\gamma_\Delta \\
y\in{\gamma_{\Delta'}}}} \Phi(|x-y|) .
\end{equation}
Taking into account the assumptions {\bf (A)} on the interaction
potential ,\,definitions  (2.1),
 (2.17)\,and the inequality\,
$|\gamma_{\Delta}|\,|\gamma_{\Delta'}|\leq\frac{1}{2}\left(|\gamma_{\Delta}|^{2}+|\gamma_{\Delta'}|^{2}\right)$\,
we obtain from \eqref{E:p_t1_1}:
\begin{equation}\label{E:p_t1_2}%(3.2)
U(\gamma)\geq
\sum_{\Delta\in\overline{\Delta_{\lambda}}:|\ga_\De|\geq
2}\;\left[E_{s}^{(N_{\Delta}(\ga))}
(\gamma_{\Delta}^{\textnormal{min}})\varphi_0-\frac{v_{0}}{2}|\gamma_{\Delta}|^{2}\right]-\frac{v_{0}}{2}|\ga|,\,
N_{\Delta}(\ga)=|\gamma_{\Delta}|.
\end{equation}
For the fixed $\varepsilon>0$\, let's define $N_{0}$\, such that
$I_{s}(\mu^{\ast};\Delta)-e_{s,\Delta}^{(N)}>\varepsilon$ if
$N<N_{0}$\\ and \,
$I_{s}(\mu^{\ast};\Delta)-e_{s,\Delta}^{(N)}<\varepsilon$ if
$N\geq N_{0}$\,(see (2.4)). Let's also define a sequence:
\begin{equation}\label{E:p_t1_3}%(3.3)
B_{N}=
  \begin{cases}
    \left(e_{s,\Delta}^{(N_{0})}-e_{s,\Delta}^{(N)}\right)\cdot
    N_{0},\, N\leq N_{0};\\
    0,\, N>N_{0}.
  \end{cases}
\end{equation}
For $N\leq N_{0}:\,e_{s,\Delta}^{(N)}-e_{s,\Delta}^{(N_{0})}\leq
0$\, and
$N^{2}\leq N\,N_{0}$. As a result we have:\\
1)\, if $N\leq N_{0}$:
\[\left(e_{s,\Delta}^{(N)}-e_{s,\Delta}^{(N_{0})} \right)N^{2}\geq
\left(e_{s,\Delta}^{(N)}-e_{s,\Delta}^{(N_{0})}
\right)N_{0}\,N=-B_{N}\,N;
\]
2)\, if $N>N_{0}$:
\[
e_{s,\Delta}^{(N)}\,N^{2}\geq e_{s,\Delta}^{(N_{0})}\,N^{2}.
\]
Then for any $N\geq2$\,:
\begin{equation}\label{E:p_t1_4}%(3.4)
 e_{s,\Delta}^{(N)}\cdot N^{2}\geq e_{s,\Delta}^{(N_{0})}\cdot
 N^{2}-B_{N}\cdot N,
\end{equation}
Because of $B_{2}>B_{N}$\, for  any $N\geq 2$\,we deduce from
(3.4)\, that for all $N\geq2$:
\begin{align}
&e_{s,\Delta}^{(N)}\,N^{2}\geq
e_{s,\Delta}^{(N_{0})}\,N^{2}-B_{2}N=\notag\\
=&I_{s}(\mu^{\ast},\Delta)N^2 +\left(e_{s,\Delta}^{(N_{0})}-
I_{s}(\mu^{\ast},\Delta)\right)\cdot N^{2}-B_{2}N\geq\notag\\
\geq &\left(I_{s}(\mu^{\ast},\Delta)-\varepsilon \right)\cdot
N^{2}-B_{2}N. \label{E:p_t1_5}%(3.5)
\end{align}
The inequality (3.5)\, proves  the Theorem 2.1 for the partition\,
$\overline{\Delta_{\lambda}}$ such that for the given
$\gamma\in\Gamma_{0}$\, there exists at least one cube with
$|\gamma_{\Delta}|\geq 2$. In this case:
\begin{equation}\label{E:p_t1_6}%(3.6)
  B=B_{2}(\varepsilon)=\left(e_{s,\Delta}^{(N_{0})}-e_{s,\Delta}^{(2)}\right)\cdot
  N_{0},\, N_{0}=N_{0}(\varepsilon).
\end{equation}
For $\gamma\in\Gamma_{0}$ with $|\gamma_{\Delta}|=1$ or 0 it is
clear that $B_{2}=v_{0}/2$. So, one can choose:
\begin{equation}\label{E:p_t1_7}%(3.7)
B=\textnormal{max}\left\{\left(e_{s,\Delta}^{(N_{0})}-e_{s,\Delta}^{(2)}\right)\cdot
  N_{0};\, \frac{v_{0}}{2} \right\}.
\end{equation}
The end of the proof.
$$
\mspace{675mu}\blacksquare
$$

\subsection{\bf Proof of Theorem 2.2 and Theorem 2.3}

In our case $K$ is a $d$-dimensional cube $\Delta$ with a rib
$\lambda$. As in the previous case we start from (3.1), (3.2). For
the fixed $\varepsilon>0$\, let us define $N_{0}$\, such that
$\left|C_{d}-\frac{E_{s}^{N}\left(\gamma_{K}^{min}\right)}{N^{2}\ln
N}\right|>\varepsilon$ if $N<N_{0}$ and \,
$\left|C_{d}-\frac{E_{s}^{N}\left(\gamma_{K}^{min}\right)}{N^{2}\ln
N}\right|<\varepsilon$ if $N\geq N_{0}$( the constant $C_{d}$ is
taken from (2.10)).\, Using (3.1), (3.2), (2.10) and neglecting in
(3.1) the part of interaction energy
$\underset{\substack{\Delta\in\overline{\Delta_{\lambda}},\\|\gamma_{\Delta}|<N_{0}}}{\sum}\;U(\gamma_{\Delta})$\,
 one can write an estimate for the
total energy of the system in the following form:
\begin{align}%(3.12)
& U(\gamma)\geq
\sum_{\substack{\Delta\in\overline{\Delta_{\lambda}},\\|\gamma_{\Delta}|\geq
2 }}\;\left[C_{d}\,\ln
|\gamma_{\Delta}|\,-\frac{v_{0}}{2}-\varepsilon
\,\ln\,|\gamma_{\Delta}| \right]|\gamma_{\Delta}|^{2}-
\sum_{\substack{\Delta\in\overline{\Delta_{\lambda}},\\|\gamma_{\Delta}|=1
}}\,\frac{v_{0}}{2}|\gamma_{\Delta}|^{2}-\notag\\
&-\sum_{i=2}^{N_{0}-1}\,\sum_{\substack{\Delta\in\overline{\Delta_{\lambda}},\\|\gamma_{\Delta}|=
i }}\,\left[C_{d}\,-\varepsilon\right]|\gamma_{\Delta}|^{2}\ln
|\gamma_{\Delta}|. \label{E:p_t2_1}
\end{align}
Number of cubes with $|\gamma_{\Delta}|=i$ is not more than
$\frac{|\gamma|}{i}$.\, That's why:
\begin{equation}\label{E:p_t2_2}%(3.13)
\sum_{i=2}^{N_{0}-1}\,\sum_{\substack{\Delta\in\overline{\Delta_{\lambda}},\\|\gamma_{\Delta}|=
i }}\,\left[C_{d}\,\,-\varepsilon\right]|\gamma_{\Delta}|^{2}\ln
|\gamma_{\Delta}|\,\leq
|\gamma|\sum_{i=2}^{N_{0}-1}\,\frac{\left[C_{d}\,\,-\varepsilon\right]
i^2\ln\, i}{i}.
\end{equation}
As a result, we can put:
\begin{equation}\label{E:p_t2_3}%(3.14)
 B=\frac{v_{0}}{2}+\sum_{i=2}^{N_{0}-1}\,\left[C_{d}\,-\varepsilon\right]i\ln\,i.
\end{equation}
The end of the proof.
$$
\mspace{675mu}\blacksquare
$$
\begin{remark}\label{R:t_third}
The proof of  the Theorem 2.3\, is very similar to the previous
proof of the Theorem 2.2. In this case according to (2.14) the
minimal energy of $N_\De(\ga)$ particles which are situated in the
$d$-dimensional cube $\De\in\overline{\De}_\la$  can be estimated
from below by the inequality (2.12)(see [3] and [10]).
Substituting this inequality in (3.2) we obtain Eq.2.21 with
$B=\frac{v_{0}}{2}$.
\end{remark}

\begin{center}{\bf Acknowledgments}\end{center}
We are grateful to Prof. Dr. V.~A.~Zagrebnov  and
Dr.~D.~L.~Finkelshtein for reading the manuscript and valuable
remarks.

\end{document}